\documentclass[journal=ancac3,manuscript=article,reprint,superscriptaddress,amsmath,amssymb,maxauthors=20]{achemso}
\usepackage{graphicx}
\usepackage[space]{grffile}
\usepackage{dcolumn}
\usepackage{bm}
\usepackage{subfigure}
\usepackage{mathrsfs}
\usepackage{booktabs}
\usepackage{notoccite}
\usepackage{float}
\usepackage{placeins}
\usepackage{tabularx}
\usepackage{color}
\usepackage{soul}
\usepackage{amsmath}
\usepackage{chngcntr}
\usepackage{multirow}

\usepackage[version=3]{mhchem} 

\author{Weinan Chen}
\email{weinanchen@psu.edu}
\altaffiliation{These authors contributed equally.}
\affiliation{Department of Materials Science and Engineering, The Pennsylvania State University, University Park, PA 16802, USA} 
\affiliation{Materials Research Institute, The Pennsylvania State University, University Park, PA 16802, USA}

\author{Disha Talreja}
\altaffiliation{These authors contributed equally.}
\affiliation{Department of Materials Science and Engineering, The Pennsylvania State University, University Park, PA 16802, USA} 
\affiliation{Materials Research Institute, The Pennsylvania State University, University Park, PA 16802, USA}

\author{Devon Eichfeld}
\affiliation{Department of Mechanical and Nuclear Engineering, The Pennsylvania State University, University Park, PA 16802, USA} 
\affiliation{Materials Research Institute, The Pennsylvania State University, University Park, PA 16802, USA}

\author{Pratibha Mahale}
\affiliation{Department of Chemistry, The Pennsylvania State University, University Park, PA 16802, USA} 

\author{Nabila Nabi Nova}
\affiliation{Department of Chemistry, The Pennsylvania State University, University Park, PA 16802, USA}

\author{Hiu Y. Cheng}
\affiliation{Department of Chemistry, The Pennsylvania State University, University Park, PA 16802, USA}

\author{Jennifer L. Russell}
\affiliation{Department of Chemistry, The Pennsylvania State University, University Park, PA 16802, USA} 
\affiliation{Department of Biochemistry and Molecular Biology, The Pennsylvania State University, University Park, PA 16802, USA}
\affiliation{Department of Physics, The Pennsylvania State University, University Park, PA 16802, USA} 

\author{Shih-Ying Yu}
\affiliation{Department of Materials Science and Engineering, The Pennsylvania State University, University Park, PA 16802, USA} 
\affiliation{Materials Research Institute, The Pennsylvania State University, University Park, PA 16802, USA}

\author{Nicolas Poilvert}
\affiliation{Department of Materials Science and Engineering, The Pennsylvania State University, University Park, PA 16802, USA} 
\affiliation{Materials Research Institute, The Pennsylvania State University, University Park, PA 16802, USA}

\author{Gerald Mahan}
\affiliation{Department of Physics, The Pennsylvania State University, University Park, PA 16802, USA} 

\author{Suzanne E. Mohney}
\affiliation{Department of Materials Science and Engineering, The Pennsylvania State University, University Park, PA 16802, USA}
\affiliation{Materials Research Institute, The Pennsylvania State University, University Park, PA 16802, USA}

\author{Vincent H. Crespi}
\affiliation{Department of Physics, The Pennsylvania State University, University Park, PA 16802, USA} 
\affiliation{Department of Materials Science and Engineering, The Pennsylvania State University, University Park, PA 16802, USA}
\affiliation{Department of Chemistry, The Pennsylvania State University, University Park, PA 16802, USA} 

\author{Thomas E Mallouk}
\affiliation{Department of Chemistry, The Pennsylvania State University, University Park, PA 16802, USA} 
\affiliation{Department of Biochemistry and Molecular Biology, The Pennsylvania State University, University Park, PA 16802, USA}
\affiliation{Department of Physics, The Pennsylvania State University, University Park, PA 16802, USA} 

\author{John V. Badding}
\affiliation{Department of Chemistry, The Pennsylvania State University, University Park, PA 16802, USA}
\affiliation{Department of Physics, The Pennsylvania State University, University Park, PA 16802, USA} 
\affiliation{Department of Materials Science and Engineering, The Pennsylvania State University, University Park, PA 16802, USA}

\author{Brian Foley}
\affiliation{Department of Mechanical and Nuclear Engineering, The Pennsylvania State University, University Park, PA 16802, USA} 
\affiliation{Materials Research Institute, The Pennsylvania State University, University Park, PA 16802, USA}

\author{Venkatraman Gopalan}
\affiliation{Department of Materials Science and Engineering, The Pennsylvania State University, University Park, PA 16802, USA}
\affiliation{Materials Research Institute, The Pennsylvania State University, University Park, PA 16802, USA}
\affiliation{Department of Physics, The Pennsylvania State University, University Park, PA 16802, USA}

\author{Ismaila Dabo}
\affiliation{Department of Materials Science and Engineering, The Pennsylvania State University, University Park, PA 16802, USA} 
\affiliation{Materials Research Institute, The Pennsylvania State University, University Park, PA 16802, USA}

\title{Achieving Minimal Heat Conductivity by Ballistic Confinement in Phononic Metalattices}


\begin{document}

\section{Abstract}
Controlling the thermal conductivity of semiconductors is of practical interest in optimizing the performance of  thermoelectric and phononic devices.  The insertion of inclusions of nanometer size in a semiconductor is an effective means of achieving such control; it has been proposed that the thermal conductivity of silicon could be reduced to 1 W/m/K using this approach and that a minimum in the heat conductivity would be reached for some optimal size of the inclusions. Yet the practical verification of this design rule has been limited. In this work, we address this question by studying the thermal properties of silicon metalattices that consist of a periodic distribution of spherical inclusions with radii from 7 to 30 nm, embedded into silicon.  Experimental measurements confirm that the thermal conductivity of silicon metalattices is as low as 1 W/m/K for silica inclusions, and that this value can be further reduced to 0.16 W/m/K for silicon metalattices with empty pores. A detailed model of ballistic phonon transport suggests that this thermal conductivity is close to the lowest achievable by tuning the radius and spacing of the periodic inhomogeneities. This study is a significant step in elucidating  the scaling laws that dictate ballistic heat transport at the nanoscale in silicon and other semiconductors.

\section{Keywords}
\textit{ ballistic heat transport, Casimir limit, metalattice,  phononic crystal, thermal conductivity, silicon}


\vspace{1 cm}

Ballistic  transport occurs when the propagating waves do not diffract among themselves or against local defects, and as such is crucial to the performance of thermocrystal waveguides \cite{anufriev2017heat}, the efficiency of thermoelectrics \cite{Schierning2014,Boukai2008a,hochbaum2008enhanced}, the reliability of electronic devices that operate under cryogenic conditions \cite{schleeh2015phonon} and the development of thermal diodes \cite{RevModPhys.84.1045, ihlefeld2015room, foley2018voltage,wehmeyer2017thermal}. This regime of thermal transport is prevalent when the length scale $\mathscr{L}$ associated with the confining network is shorter than the phonon mean free path $\ell$. For a typical semiconductor of lattice heat conductivity $\kappa_0$, the mean free path can be estimated as $\ell \propto (\kappa_0 V)/(N v k_{\rm B})$ (where $k_{\rm B}$ is the Boltzmann constant, $v$ denotes the sound velocity, and $V$ is the volume of the unit cell comprising $N$ atoms), yielding  $\ell \sim$ 50-1,000 nm \cite{feng2014prediction}. Hence, a transition to the ballistic regime takes place when the length $\mathscr{L}$ describing geometric confinement is brought down to a few tens of nanometers \cite{Lee2015, romano2016temperature, chiu2005ballistic, chen1998thermal, anufriev2017heat, maire2017ballistic}.   

\begin{figure}
    \centering\includegraphics[width=0.8\columnwidth]{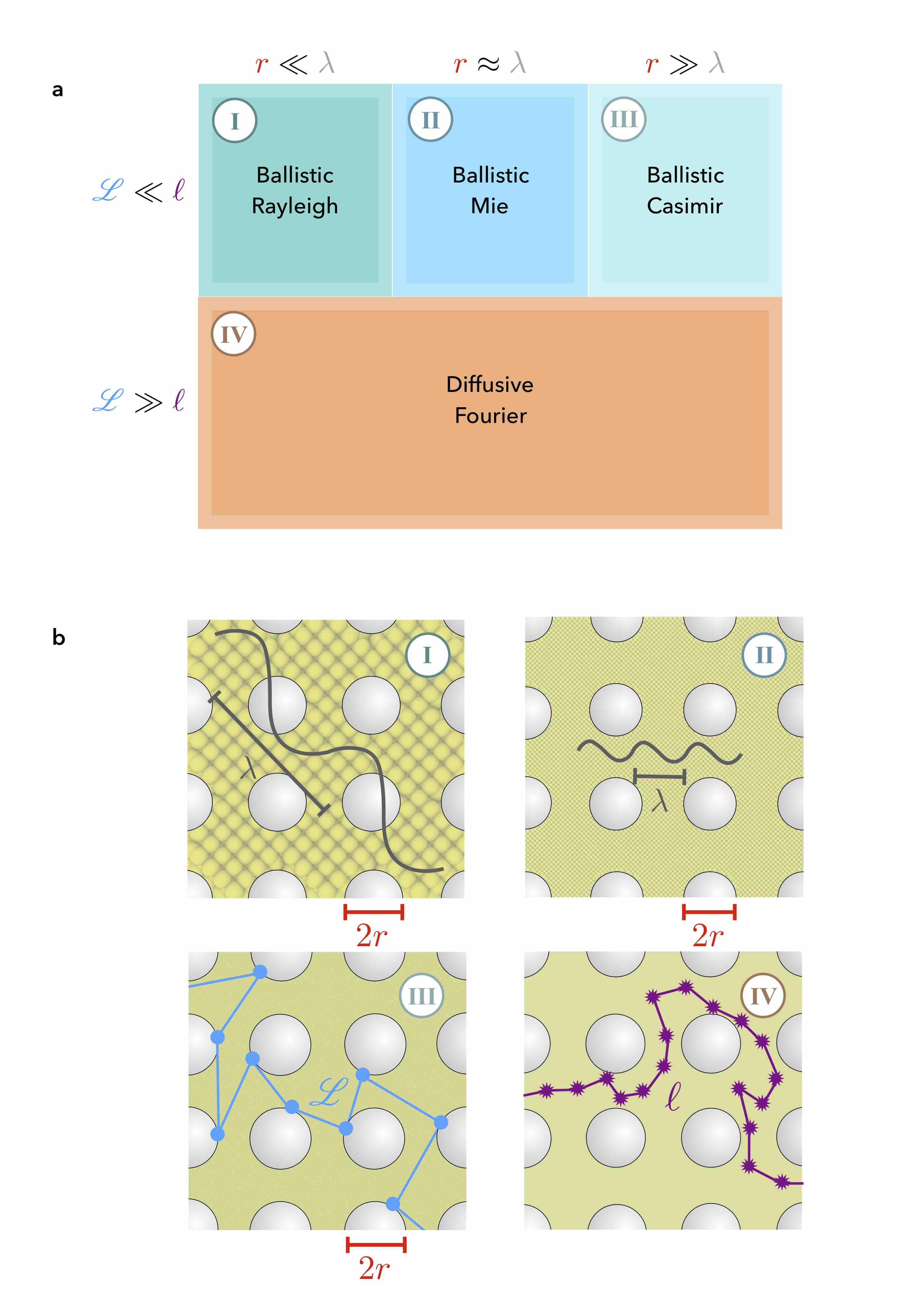}
    \caption{\small  (a) Regimes of phonon propagation across a network of spherical inclusions. The variables $r$, $\lambda$, $\ell$, and $\mathscr{L}$ denote the radius of the cavity, the wavelength of the incident phonon, the intrinsic phonon mean free path and the phonon ballistic length, respectively. (b) Illustrations of the three ballistic regimes and the conventional diffusive limit. The circular deflection points represent phonon-boundary collisions. The stars depict phonon-phonon scattering events.}    
    \label{fig:phenomenological-mapping}
\end{figure}

It has been postulated that the dispersion of spherical inclusions smaller than 100 nm in a semiconducting matrix can lead to a thermal conductivity minimum. As shown in Fig.~\ref{fig:phenomenological-mapping},  in the Rayleigh limit where the radius $r$ of the spherical inhomogeneities vanishes (Regime {\sc i}), the thermal conductivity $\kappa$  decreases rapidly with $r$ according to the scaling law $(\kappa_{\rm 0} - \kappa)/\kappa_{\rm 0} \propto \ell/\mathscr{L} \propto r^3 \ell/\lambda^4$\cite{ying1956scattering}. Conversely, in the Casimir limit where the sphere size is very large in comparison to the incident wavelength (Regime {\sc iii}), the thermal conductivity increases linearly with respect to sphere size: $(\kappa_{\rm 0} - \kappa)/  \kappa_{\rm 0} \propto \ell/\mathscr{L} \propto \ell/r$. \cite{hsiao2013observation, Lee2015} At the juncture of these two  trends (Regime {\sc ii}), a minimum in the heat conductivity is expected  \cite{mingo2009nanoparticle, wang2009improved, feser2019engineering}. 

\begin{figure}
    \centering\includegraphics[width=0.5\columnwidth]{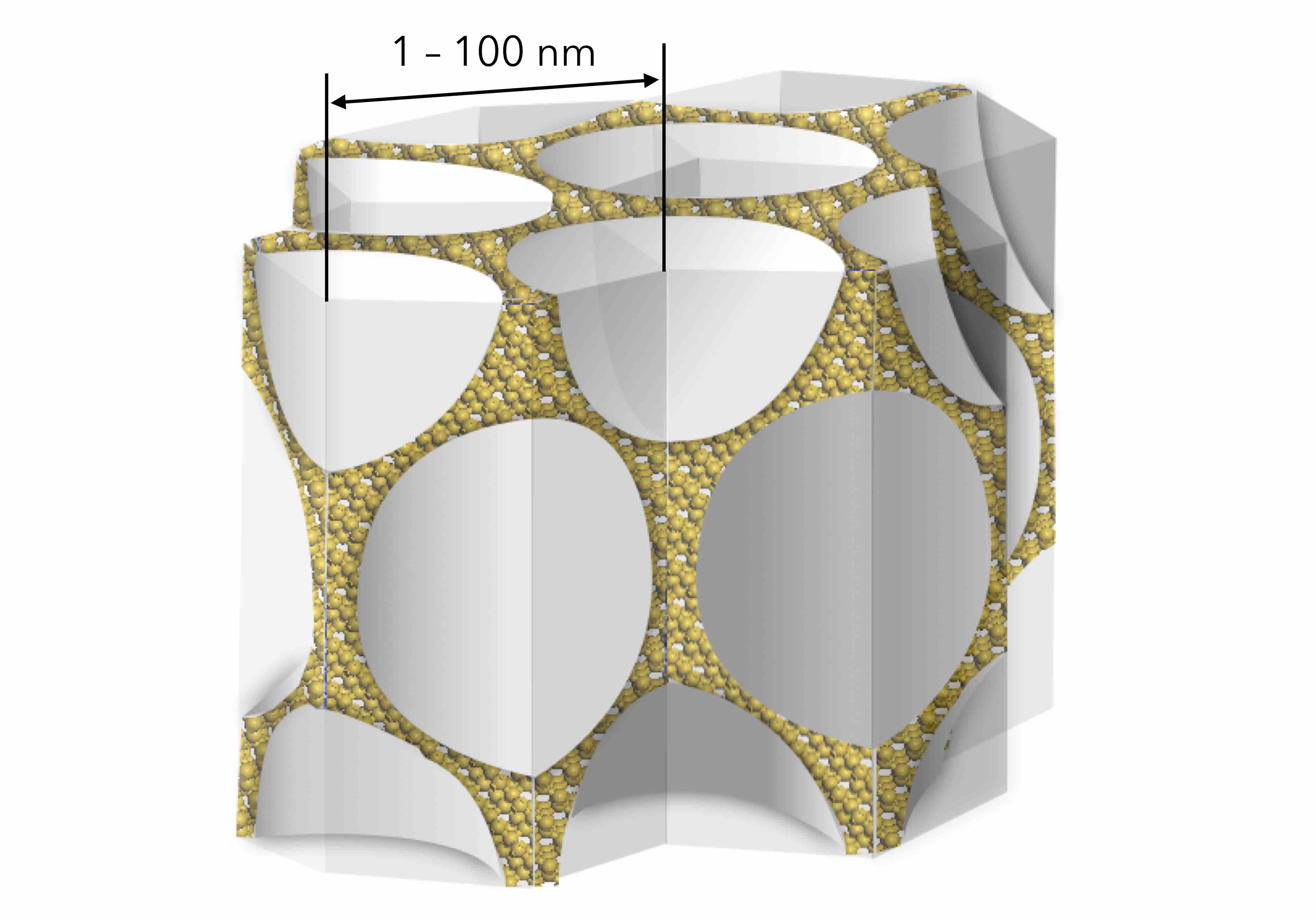}
    \caption{\small Metalattice consisting of a closely spaced distribution of spherical voids in crystalline silicon.  Tuning the diameter of the pores in the range 1-100 nm, corresponding to the mean free paths of the heat-carrying phonons, affords sensitive control on ballistic heat transport. }
    \label{fig:metalattice}
\end{figure}

In this work, we aim to examine the possibility of this thermal conductivity minimum. To this end, we synthesize and characterize a class of metamaterials consisting of a periodic arrangement of nanometer-sized spherical inclusions in a semiconducting crystal (Fig.~\ref{fig:metalattice}) \cite{Han2001a,Han2002}, which we refer to here as metalattices. Similar nanostructures have been synthesized previously, by means of conventional chemical vapor deposition, albeit on a scale that is an order of magnitude larger \cite{ma2013coherent, barako2016quasi, barako2014thermal, padmanabhan2014bottom}. In this work, silicon metalattices with sphere radii in the range of 7 to 30 nm are obtained by high-pressure infiltration and selective etching of colloidal templates \cite{liu2018confined}.   We undertake a comprehensive experimental and theoretical study of the thermal conductivity of metalattices, with a focus on verifying and explaining the occurrence of a minimal lattice thermal conductivity as a function of size, volume fraction, and the content of the spherical inclusions. 

\section{Results and discussion} 

\begin{figure}
	\centering\includegraphics[width=1\columnwidth]{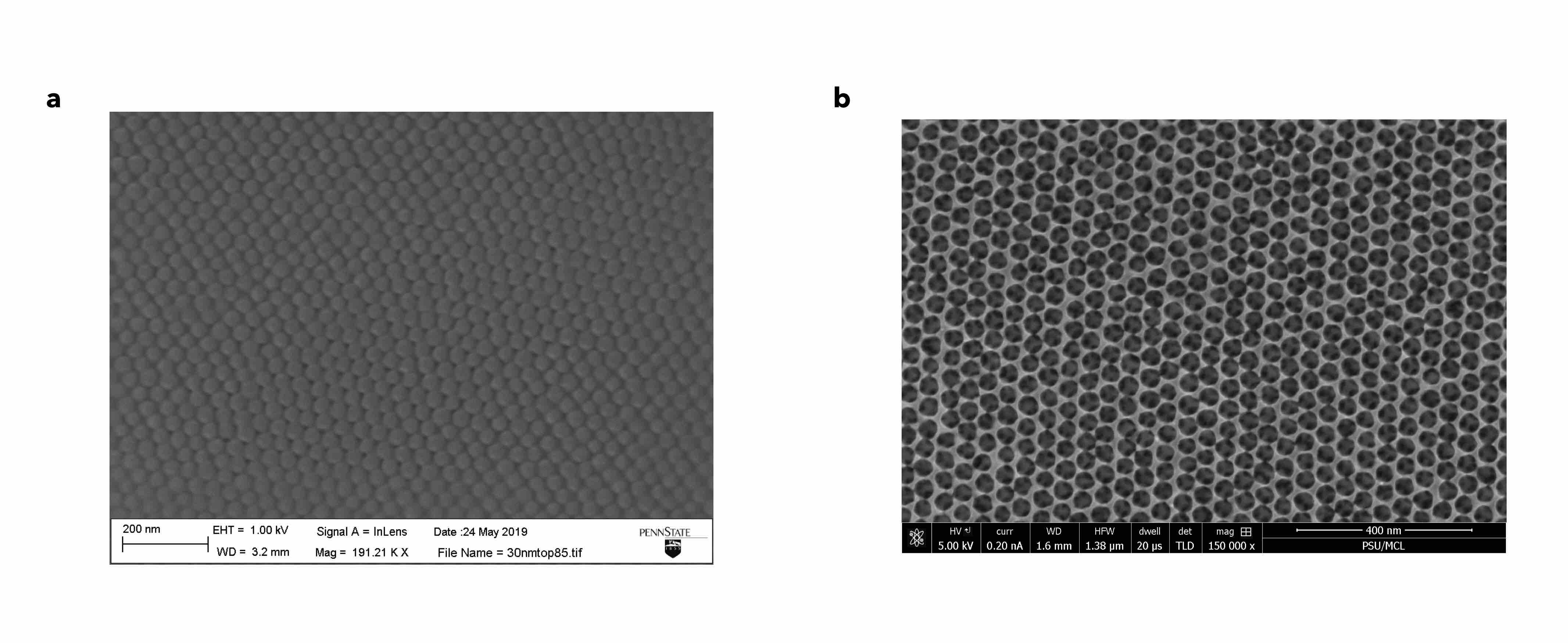}
	\caption{\small  Scanning electron micrographs of silicon metalattices with (a) filled and (b) empty spheres of average radius $r = $ 15 nm. }
	\label{fig:SEM_image}
\end{figure}

Two types of silicon metalattices are synthesized: one with empty pores and one with silica filled pores. While a number of studies have focused on synthesizing and investigating thermal transport in two-dimensional nanoporous systems \cite{Yu2010a, Tang2010, lim2016simultaneous, alaie2015thermal, maire2017heat, anufriev2017heat, verdier2017thermal}, there are limited reports on the thermal transport of three-dimensional metalattice (or inverse opal) nanostructures\cite{ma2013coherent, barako2016quasi, barako2014thermal, padmanabhan2014bottom}. It should be noted that the sizes of the spheres of previously reported structures are in the range of 100 to 1,000 nm. Here, in order to probe the size regime that potentially achieves minimal thermal conductivity,  silicon metalattices with sphere radii from 7 to 30 nm are synthesized by infiltrating the interstitial voids of an ordered template assembled from silica nanospheres by means of high-pressure confined chemical vapor deposition (HPcCVD)\cite{baril2010high, baril2011confined, sparks2013conformal}.  Using a similar method, magnetic metalattices with average sphere radius as small as 15 nm have been previously synthesized\cite{liu2018confined}.  The synthesis of silicon metalattices starts with making a colloidal solution of mono-dispersed silica nanospheres,  
using the seeding and regrowth process described by \citet{hartlen2008facile} and \citet{watanabe2011extension}. These silica spheres are subsequently assembled as closely packed templates on a silicon substrate using vertical deposition techniques \cite{kuai2004high, russell2017binary}. In this method,  after cleaning the silicon substrate using a Piranha solution, the sample is placed with an inclination angle of approximately 30 degrees in a vial that contains a colloidal suspension formed by diluting the above-prepared silica nanoparticles. Subsequently, the vial is placed in an oven where the solvent evaporates off under controlled temperature and humidity conditions, leaving behind a silica template film on the substrate. Depending on sphere size and deposition conditions,  the thicknesses of these films vary from 250 to 2,000 nm, as observed from electron microscope imaging. The template film is further calcined at 600 $\rm ^oC$ to remove the remnant organic contamination. Voids of this template film are then infiltrated with silicon using high-pressure confined chemical vapor deposition. Deposition time is adjusted for each sample such that a silicon over-layer is formed on the top of the template. This ensures complete void filling of the structures. After removing this over-layer using deep reactive ion etching, the infiltrated amorphous silicon is crystallized by annealing at 800 $\rm{^o}$C for 30 minutes under an inert atmosphere. A scanning electron microscope image of the resulting silica-filled silicon metalattice with an average sphere radius of 15 nm is shown in Fig.~\ref{fig:SEM_image}(a).  To make silicon metalattices with empty pores, the silica-filled metalattices are etched with a vapor of 49\% hydrofluoric acid. The resulting empty silicon metalattice is shown in the scanning electron microscope image of Fig.~\ref{fig:SEM_image}(b). 

Thermal conductivity measurements are performed using time domain thermoreflectance (TDTR) \cite{cahill2002thermometry, cahill2014nanoscale, hopkins2010criteria}. All measured samples are deposited with aluminum as a transducer layer using electron beam evaporation \cite{thomsen1984coherent,thomsen1986surface}, with  thicknesses ranging from 80 to 120 nm as determined using picosecond acoustics. We measure 4-5 different spots on each sample for room temperature thermal conductivity data (the experimental procedure and parameters are detailed in the \textit{Methods} section).  Low temperature thermal conductivity measurements are performed with the samples loaded inside a liquid helium continuous flow optical cryostat, providing temperature stability within 1 K.

\begin{table}
\caption{Thermal conductivity of metalattices with empty and filled pores measured by time domain thermoreflectance.}
\label{table:experimental_TDTR_results}
\begin{tabular}{c c c c c c c}
\\
\toprule \specialrule{1pt}{1pt}{1pt}
\\
\multirow{2}{*}{Type} & \multirow{2}{*}{Temperature}  & \multicolumn{4}{c}{Radius} \\\cmidrule{3-6}

&   & 7 nm & 10 nm& 15 nm & 30 nm \\ 
 \\
 \midrule \specialrule{0.5pt}{1pt}{1pt}
 \\
 Filled & 140 K  & 1.11 $\pm$ 0.08 & 0.79 $\pm$ 0.05 & 1.45 $\pm$  0.09 & 1.51 $\pm$ 0.08  & W/m/K\\
 & 200 K  & 1.26 $\pm$ 0.09 & 0.91 $\pm$ 0.06 & 1.69 $\pm$ 0.11 & 1.74 $\pm$ 0.10 \\
 & 300 K & 1.33 $\pm$ 0.09 & 0.94 $\pm$ 0.06   & 1.68 $\pm$ 0.11 & 2.08 $\pm$ 0.12 \\  
 Empty & 300  K &  0.16 $\pm$ 0.05 & 0.24 $\pm$ 0.04   & 0.26 $\pm$ 0.07 & \\
\\
\toprule \specialrule{1pt}{1pt}{1pt}
\\

\end{tabular}
\end{table}

The experimentally measured thermal conductivities are reported in Table~\ref{table:experimental_TDTR_results}.  From 140 to 300 K, the thermal conductivities  of filled metalattice samples show a minimum as a function of the sphere radius. However, such a minimum is not found in empty silicon metalattices: the thermal conductivity increases with increasing sphere sizes. On the other hand, it is observed that the thermal conductivities of empty metalattices are ten times smaller than the filled ones, with the lowest value being 0.16 W/m/K. In order to understand the influence of sphere content (empty or filled), we develop and validate a computational model based on the Casimir theory of ballistic transport. 

In this model \cite{Casimir1938}, the heat current density at an arbitrary point $\boldsymbol{r}_1$ can be expressed as 
\begin{align}
\boldsymbol{j}(\boldsymbol{r}_1) = \frac{1}{{V}N_{\boldsymbol{q}}} \sum_{\lambda\boldsymbol{q}} \int_{\mathscr{S}_2} \hbar\omega_{\lambda\boldsymbol{q}} n_{\lambda \boldsymbol{q}}(\boldsymbol{r}_2)\boldsymbol{v}_{\lambda\boldsymbol{q}} \frac{d\Omega_2}{4 \pi}
\label{eqn:casimir-current-density}
\end{align}
where $\omega$ and $\boldsymbol{v}_{\lambda\boldsymbol{q}} $ are the frequency and group velocity of the phonon with wave vector $\boldsymbol{q}$, polarization $\lambda$, and statistical occupancy $n = \left(\exp{\left(\hbar\omega/k_{\rm B} T\right)} - 1\right)^{-1}$, with the sum being carried out over a sampling of $N_{\boldsymbol{q}}$ wave vectors in reciprocal space. Here, $\mathscr{S}_2$ defines the area that is ballistically accessible from the point $\boldsymbol{r}_1$
and $\Omega_2$ is the total solid angle associated with $\mathscr{S}_2$. For small temperature gradients, Eq.~\eqref{eqn:casimir-current-density} can be recast as  
\begin{align}
\boldsymbol{j}({\boldsymbol{r}_1})  =  \frac{1}{{V}N_{\boldsymbol{q}}} \sum_{\lambda\boldsymbol{q}}& \int_{\mathscr{S}_2} c_{\lambda \boldsymbol{q}}  \boldsymbol{v}_{\lambda\boldsymbol{q}}   (\boldsymbol{r}_{12} \cdot \nabla T) \frac{d\Omega_2}{4\pi},
\end{align}
with $c = \hbar\omega \partial n / {\partial T} $ being the phonon heat capacity. The heat flux across a surface area $\mathscr{A}_1$ is then calculated as
\begin{align}
I_{1}  = \frac{1}{{V}N_{\boldsymbol{q}}} \sum_{\lambda\boldsymbol{q}} \int_{\mathscr{A}_1}\int_{\mathscr{S}_2}  c_{\lambda \boldsymbol{q}} v_{\lambda \boldsymbol{q}} |\nabla T|  (\boldsymbol{r}_{12} \cdot \hat{\boldsymbol{e}}_{\parallel})( \hat{\boldsymbol{r}}_{12} \cdot d\boldsymbol{\sigma}_1)   \frac{d\Omega_2}{4\pi}.
\end{align}
Here, $\hat{\boldsymbol{e}}_{\parallel} = - \nabla T/|\nabla T|$ is a unit vector antiparallel to the temperature gradient (and where use has been made of the fact that $\hat{\boldsymbol{v}}_{\lambda \boldsymbol{q}} = - \hat{\boldsymbol{r}}_{12}$). Since the temperature gradient is small, the change in the heat capacity from the emitting surface to the receiving area is negligible, so that both the heat capacity and the group velocity of the phonon can be taken out of the integral. We thus obtain an expression for the heat conductivity that is reminiscent of the solution to the linearized Boltzmann transport equation:
\begin{align}\label{eqn:kappa_final}
\kappa  =  \frac{1}{3{V}N_{\boldsymbol{q}}} \sum_{\lambda\boldsymbol{q}}  c_{\lambda\boldsymbol{q}}  v_{\lambda\boldsymbol{q}} \mathscr{L},
\end{align}
where the ballistic length reads
\begin{align}\label{eqn:Casimir_length}
 \mathscr{L} = \frac{1}{\mathscr{A}_1}\int_{\mathscr{A}_1} \mathcal{L}(\boldsymbol{r}_1)d\sigma_1
\end{align}
with
\begin{equation}{\label{eqn:L_r1}}
\mathcal{L}(\boldsymbol{r}_1) = \frac{3}{4\pi} \int_{\mathscr{S}_2}  (\boldsymbol{r}_{12} \cdot \boldsymbol{e}_{\parallel}) \cos{\theta_1}d\Omega_2.
\end{equation}

\begin{figure}
	\centering\includegraphics[width=0.8\columnwidth]{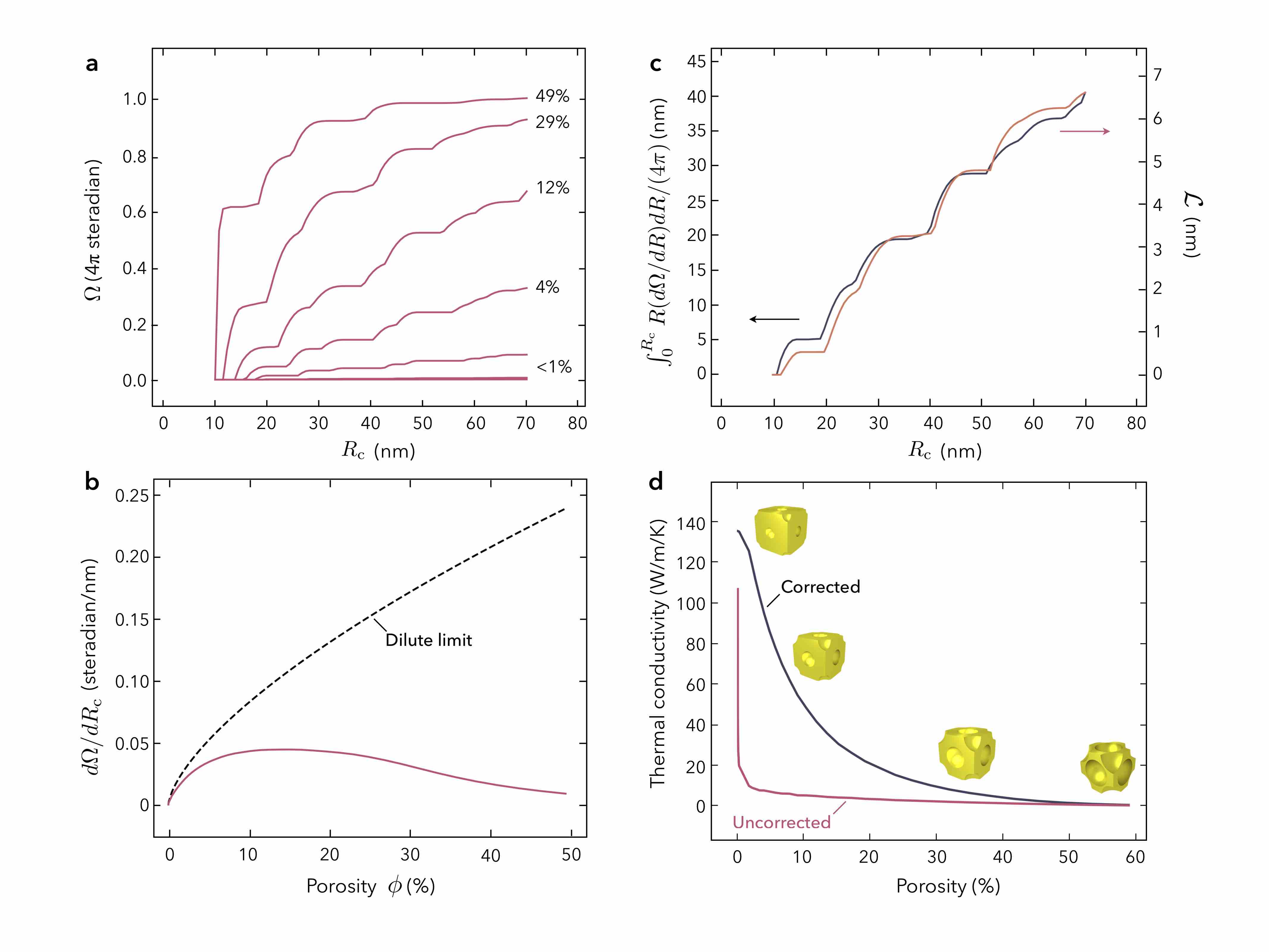}
	\caption{\small Long-range correction to the ballistic length. (a) Total solid angle captured in a spherically truncated domain as a function of the cutoff radius $R_{\rm c}$ and porosity $\phi$. The system is a simple cubic metalattice with a fixed period ($p$ = 20 nm). (b) Slope $d\Omega/dR_{\rm c}$ of the solid angle as a function of the cutoff radius at ${R_{\rm c}}$ = 70 nm (solid line). The dashed line represents the low-porosity asymptotic approximation $ \sim$$(6\pi^2\phi)^{2/3}/p$.  (c) Comparison of the ballistic length $\mathcal{L}$ with its radial approximation $\int_{0}^{R_{\rm c}} R ({d\Omega}/{dR_{\rm c}}) dR/(4\pi)$. $\mathcal{L}$ and  $\int_{0}^{R_{\rm c}} R ({d\Omega}/{dR_{\rm c}}) dR/(4\pi)$ are in quantitative agreement up to a proportionality constant $\eta$. (d) Comparison of the thermal conductivity predictions with and without solid angle correction.  }
	 \label{fig:L_correction}	 
\end{figure}

The calculation of the  mean free path $\mathscr{L}$ necessitates a simulation domain of finite size. Here, we adopt a spherical truncation of cutoff radius $R_{\rm c}$. Figure~\ref{fig:L_correction} examines the influence of this truncation on the convergence of the calculated ballistic length $\mathscr{L}$. As one might expect, when the porosity $\phi$ is small, the total solid angle captured inside the simulation domain is significantly lower than $4\pi$ steradians, as shown in Fig.~\ref{fig:L_correction}(a). To restore the omitted contribution, we develop a model that consists of including the solid angle $\Omega$ beyond $R = R_{\rm c}$ by estimating the distance $R_{\rm m}$ at which the solid angle reaches $4\pi$ steradians $R_{\rm m}=R_{\rm c}+[{4\pi-\Omega(R_{\rm_c})}] / [{{d\Omega}(R_{\rm c})/{d R}}]
$. The slope of the solid angle at $R = R_{\rm c}$ is shown in Fig.~\ref{fig:L_correction}(b). In the dilute regime where the pores are small and well separated, each pore contributes the same amount of solid angle (neglecting shadowing effect). In this limit, the slope $d\Omega / d R$ can be expressed as $\sim$$(6\pi^2\phi)^{2/3}/p$, where $p$ is the period of the metalattice. It is then possible to obtain the corrected ballistic length scale $\mathcal{L}(\boldsymbol{r}_1)$ from the uncorrected value $\mathcal{L}_{\rm c}(\boldsymbol{r}_1)$:

\begin{equation}\label{eqn:final_correction}
\mathcal{L}(\boldsymbol{r}_1) = \mathcal{L}_{\rm c}(\boldsymbol{r}_1) + \frac {R_{\rm m}^2 - R_{\rm c}^2}{8\pi\eta} \frac{d\Omega}{d R}(R_{\rm c}),
\end{equation}
where $\eta$ is the geometric factor that relates $\mathcal{L}_{\rm c}$ to its angular-averaged approximation, as shown in Fig.~\ref{fig:L_correction}(c). The thermal conductivities of metalattices calculated with and without long-range correction are compared in Fig.~\ref{fig:L_correction}(d), showing the importance of this corrective procedure in computing the ballistic length scales at low porosity. The procedure we proposed can be used to study thermal transport in the dilute limit in other phononic systems such as two-dimensional phononic nanostructures, and can be implemented easily within other numerical approaches such as Monte Carlo calculations \cite{hua2017anisotropic, hua2017efficient, liao2018akhiezer, bera2010marked}. 

We now apply this model to determine the thermal conductivity of metalattices.  The vibrational spectra $\omega_{\lambda\boldsymbol{q}}$ and phonon-phonon mean free paths $\ell_{\lambda\boldsymbol{q}}$ are derived from semilocal density-functional theory. The effect of Rayleigh scattering on ballistic transport is included through the rescaling $\mathscr{L}\to \mathscr{L}/\gamma_{\lambda}$, in which $\gamma_{\lambda}$ describes the subwavelength enhancement of the ballistic cross section (the scattering efficiency) \cite{kraft1971scattering,ying1956scattering}. The calculations of the scattering efficiency, the effect of Brillouin zone folding on phonon group velocity, as well as the effect of phonon-isotope scattering are discussed in the \textit{Supplementary Information}. 

Figure \ref{fig:thermal_conductivity_temperature_dependence}(a) reports the thermal conductivities of filled metalattices as a function of porosity and temperature. In the low porosity limit (1\%), the thermal conductivity decreases with increasing temperature, a behavior that is reminiscent of bulk silicon. However, in the high-porosity limit (70\%), the thermal conductivity shows a trend that is characteristic of ballistic phonon transport\cite{romano2016temperature, maire2017ballistic, chiu2005ballistic,chen1998thermal,anufriev2017heat,Lee2015}; $\kappa$  increases gradually as the temperature is raised from 50 to 250 K, before reaching a maximum. Experimental measurements on samples with the same geometric parameters exhibit a similar saturation [Fig.~\ref{fig:thermal_conductivity_temperature_dependence}(b)], thereby providing clear evidence that ballistic transport is prevalent in silicon metalattices. This trend has been previously predicted in two-dimensional metamaterials. \cite{romano2016temperature} Compared to the two-dimensional case, the thermal conductivities of metalattices are reduced to much lower values due to a sharper decrease in the scattering cross section in higher dimension. Moreover,  while the asymptotic values of thermal conductivity at room temperature are relatively well predicted by the ballistic model, there is a larger discrepancy between theory and experiment at  lower temperature (50 to 150 K) for the empty metalattices. This discrepancy is likely due to specular scattering, which causes an increase in the phonon ballistic mean free paths.\cite{srivastava1990physics} This effect is more pronounced for long wavelength phonons that dominate heat transport at low temperatures.\cite{ravichandran2018spectrally, lindsay2018survey}
\begin{figure*}
	\centering\includegraphics[width=1\columnwidth]{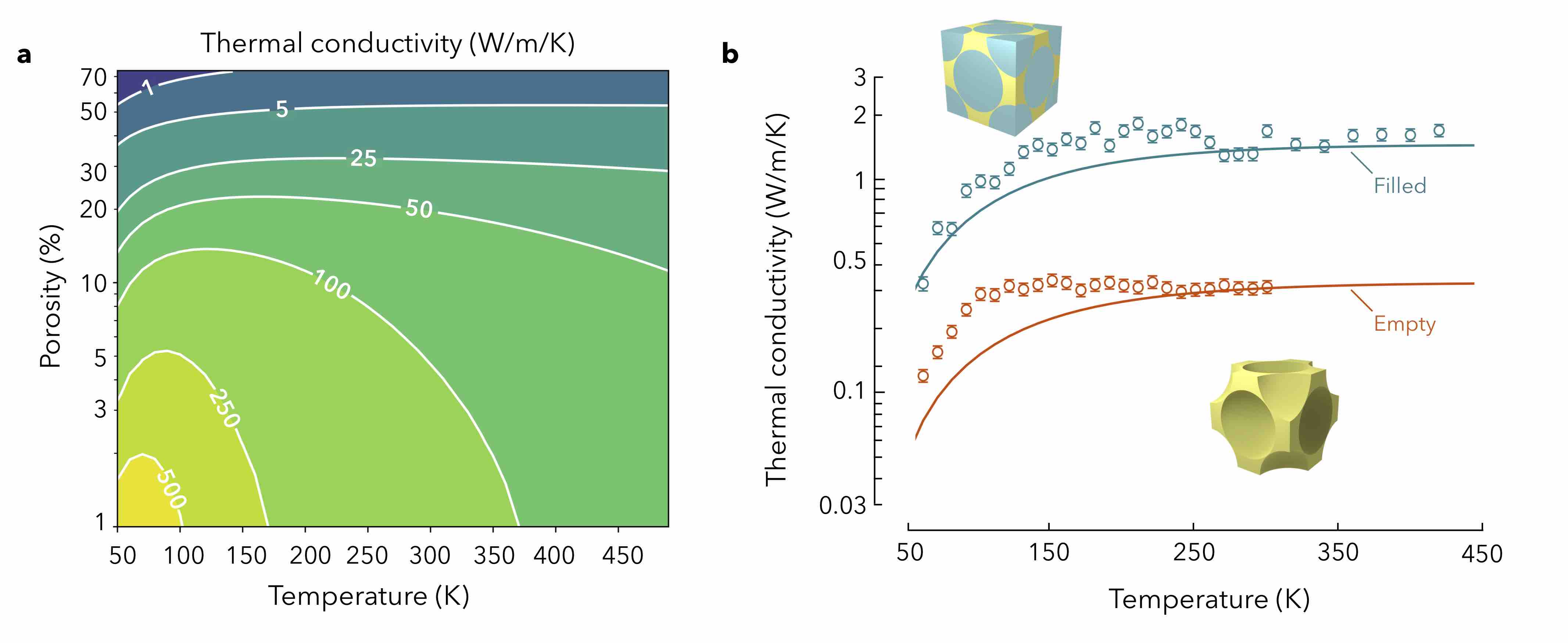}
	\caption{\small  (a) Thermal conductivity of silicon metalattices with silica-filled spheres as a function of porosity and temperature. The pore radius is fixed at 15 nm.  (b) Comparison between experimental and theoretical data for the temperature-dependent thermal conductivities of face-centered cubic close-packed metalattices with a pore radius of 15 nm.  }
	\label{fig:thermal_conductivity_temperature_dependence}
\end{figure*}

\begin{figure*}
	\centering\includegraphics[width=1\columnwidth]{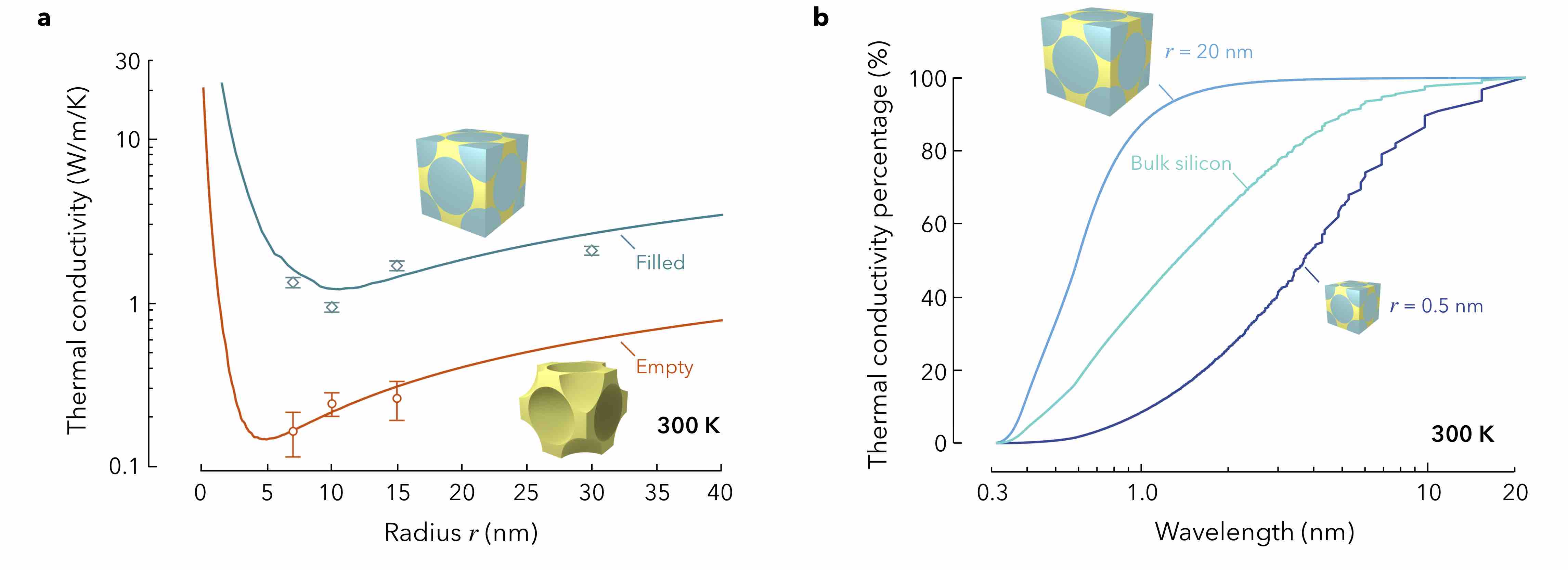}
	\caption{\small (a) Thermal conductivity of silicon metalattices as a function of sphere radius with filled and empty pores. (b)  Comparison between the wavelength-decomposed thermal conductivity of silicon metalattices and bulk silicon. Two metalattices with face-centered cubic packing and the same porosity (73\%), but different pore radii (0.5 nm and 20 nm), are considered. }
	\label{fig:thermal_conductivity_radius_dependence}
\end{figure*}

In Fig.~\ref{fig:thermal_conductivity_radius_dependence}(a), we investigate the thermal conductivities of metalattices as a function of pore radius. Our model reveals that the occurrence of the thermal conductivity minimum depends on the content of the inclusions: for silicon metalattices with empty and silica-filled spheres, simulations predict a thermal conductivity minimum at a composition-dependent radius (5 nm for empty metalattices and 10 nm for filled metalattices).  Experimental measurements strongly suggest the existence of a minimal heat conductivity in filled silicon metalattices, but not in empty metalattices. Our simulation points out that  an even smaller sphere radius is needed to reach the minimal thermal conductivity in empty metalattices. The thermal conductivity predictions from our model are in quantitative agreement with experimental measurements, implying that ballistic transport is likely responsible for the considerably reduced  conductivity of silicon metalattices. We note that the current implementation of the ballistic model is based on finite-element discretization, where the surfaces of the pores are kept rigid. Investigating the effect of local surface distortion  could potentially increase the reliability of the present calculations.  Moreover, we note that the theoretical predictions are based on perfect metalattice structures, while there exist small deviations in the size, shape and dispersity of the pores experimentally. Taking these effects into consideration is a relevant topic for further theoretical development.

In Fig.~\ref{fig:thermal_conductivity_radius_dependence}(b), we examine the different ballistic regimes by decomposing the thermal conductivity of metalattices onto a wavelength spectrum. We consider two face-centered cubic metalattices with different radii, namely, 0.5 nm and 20 nm. The wavelength-decomposed thermal conductivity of bulk silicon is also plotted for reference. It is shown that the two metalattices show distinct behaviors, \textit{i.e.}, phonons that mainly contribute to the thermal conductance in metalattices with  small pores (0.5 nm) have long wavelengths due to Rayleigh scattering,  whereas 90\% of heat propagation through metalattices with large pore size (20 nm) is facilitated by phonons with extremely short wavelengths (smaller than 1 nm), corresponding to the Casimir regime. Figure~\ref{fig:thermal_conductivity_radius_dependence} demonstrates the possibility of adjusting the sphere sizes to filter out specific wavelengths and achieving sensitive control over thermal transport. 

In closing, we discuss the practical significance of the observed thermal conductivity minimum. Potential applications of the geometry-dependent thermal conductivity of phononic metalattices include the design of thermal rectifiers. By incorporating inclusions of different sizes across a semiconductor, different regions of the material could have  different temperature-dependent thermal conductivities. This spatial gradient of pore radii will result in a drastically different effective thermal resistance depending on the direction of the temperature bias. The thermal rectification performance of graded metalattices using various semiconductors is currently being studied  and could ultimately enable the realization of efficient thermal diodes.

\section{Conclusion}
\label{sec:conclusion}
We have investigated thermal transport in silicon metalattices to verify the existence of a thermal conductivity minimum  as a function of the radius of the spherical inclusions. Thermal conductivity measurements in silica-filled silicon metalattices strongly imply the occurrence of this minimum when the radius of the pore is brought down to 10 nm in the close packing limit. The thermal conductivity of empty metalattices is shown to be as small as 0.16 W/m/K for  a sphere radius of 7 nm. Using a well-converged phonon ballistic model, we have found indications that the thermal conductivity may be further lowered by reducing the sphere radius to 5 nm. This work highlights the potential of metalattices as a widely applicable platform to minimize the thermal conductivity of crystalline semiconductors, motivating further work to push the frontier of nanofabrication into the Rayleigh limit of heat transport.  
\section{Methods}
\label{sec:methods}
\appendix
\renewcommand\thefigure{\thesection.\arabic{figure}} 
\setcounter{figure}{0}  

\textit{Finite-element simulations.} Finite-element models are constructed to compute ballistic  mean free paths $\mathscr{L}$ using  {\sc python} and the {\sc trimesh} module \cite{trimesh}. Spherical inclusions are modeled by two-dimensional surface triangular meshes, and a ray-casting algorithm is employed to find the accessible areas for ballistic propagation. Each spherical surface mesh contains 2,068 triangular elements. The cutoff radius $R_{\rm m}$ is chosen so that at least 200 explicit spheres are included in the simulation. The longitudinal and transverse scattering efficiencies are computed according to \citet{kraft1971scattering, ying1956scattering}.  The density, Poisson's ratio, and Young's modulus are taken to be $\rho = 2,330 \,\rm {kg/m^3}$, $\nu = 0.27$, $E = 127 $ GPa for silicon, and $2,230 \,\rm {kg/m^3}$, $ 0.17$, $ 73.1$ GPa for silica.

\textit{First-principles simulations.} The force constant matrix is obtained from density-functional theory calculations using a 3 $\times$ 3 $\times$ 3 $\rm{\boldsymbol{q}}$ point mesh. Self-consistent calculations are done using the Perdew-Burke-Ernzehof approximation with the {\sc pw} code of the Quantum-Espresso software package \cite{perdew1996generalized,giannozzi2009quantum}. Norm-conserving pseudopotentials are used with kinetic energy cutoffs of 100 and 400 Ry. The Brillouin zone is sampled with a 3 $\times$ 3 $\times$ 3 Monkhorst-Pack grid \cite{marzari1999thermal}.  Phonon relaxation times are computed using the {\sc phono3py} package based on the computed harmonic and anharmonic force constants \cite{phono3py}.  The resulting bulk thermal conductivity of silicon at 300 K using a $48 \times 48 \times 48$ $\boldsymbol q$-point extrapolation is 135 W/m/K, in good agreement with the experimental value \cite{kremer2004thermal}.

\textit{Time-domain thermoreflectance measurements.} Time-domain thermoreflectance (TDTR) is applied to measure the thermal conductivity of the silicon metalattices. A titanium-sapphire laser centered at 800 nm wavelength and emitting 100 fs pulses at a repetition rate of 80 MHz is used. The emitted beam is divided into pump and probe beam paths using a polarizing beam splitter with a half waveplate controlling the power distribution in the two beam paths. The pump beam is modulated by a square wave at a frequency $\nu_p = 8.75 $ MHz using an electro-optic modulator, which serves as the reference frequency for signal collection by the lock-in amplifier. It is then passed through a bismuth borate crystal to convert its wavelength from 800 nm to 400 nm using second harmonic generation. Color filters are placed before the detector to help prevent any spurious signal contribution from the scattered pump beam at the modulation frequency. The probe beam is sent through a delay stage that can provide a maximum time delay of 6 ns with respect to the pump beam. Both beams are then coaxially focused through a 5$\times$-magnification microscope objective onto the sample surface. The focal spot sizes used in the system are 21 $\mu$m for the pump beam and 5-10 $\mu$m for the probe beam as measured by knife edge technique. A complementary metal-oxide semiconductor (CMOS) camera is used in this setup to create a dark field image of the sample surface during each measurement.
The pump beam absorbed by aluminum on the sample surface creates a modulated surface heating, and the temperature decay is monitored by a time-delayed probe beam reflectivity. The reflected beam is focused through a lens onto a silicon photodiode detector. It is connected to a wide (RF) lock-in amplifier \textit{via} a resonant (RLC) circuit which collects the in-phase component $V_{\rm in}$ and out-of-phase component $ V_{\rm out}$ of the thermoreflectance signal as a function of delay time at a resonance corresponding to the modulation frequency. The schematic describing the experimental setup for the time-domain thermoreflectance measurement is available in the \textit{Supplementary Information}.

We measure at least 4 different spots on each sample for room temperature thermal conductivity data. While fitting the experimental ratio $- V_{\rm in}/ V_{\rm out}$ to the thermal model,  eight unknown parameters, namely, the thermal conductivity of the silicon substrate ($\Lambda_{\rm Si}$),  volumetric heat capacity of the silicon substrate ($C_{\rm Si}$), thermal conductivity of aluminum ($\Lambda_{\rm Al}$), volumetric heat capacity of aluminum ($C_{\rm Al}$), interfacial thermal conductance between aluminum and the metalattice film ($G_{ \rm Al/film}$), and interfacial thermal conductance between the metalattice film and the silicon substrate ($G_{\rm film/Si}$), thermal conductivity of the film ($\Lambda_{\rm film}$), and volumetric heat capacity of the film ($C_{\rm film}$) need to be determined. Amongst these, literature reported values are taken for the thermal conductivity and volumetric heat capacity of aluminum and silicon, respectively. \cite{touloukian1970thermophysical} The heat capacity of the metalattice film is calculated using the rule of mixtures $C_{\rm film} = (1 - \phi) C_{\rm Si} + \phi\, C_{\rm pore}$, where $\phi$ is the volume fraction of the pores in the metalattice film and $C_{\rm pore}$ and $C_{\rm Si}$ are the volumetric heat capacities of  pore material and silicon.  Thus, for filled metalattices, $C_{\rm pore}$ is taken as the literature reported value for the volumetric heat capacity of silica. For empty metalattices, the first term is neglected due to the negligible volumetric heat capacity of air with respect to silicon.  The sensitivity analysis of experimental parameters, as well as the original room temperature time-domain thermoreflectance measurement data for empty and filled metalattices at different sphere sizes, are included in the \textit{Supplementary Information}. 

The thermal penetration depth in the samples is estimated from 
\begin{equation}
d = \sqrt{{\Lambda_{\rm film}}/(\pi C_{\rm film}\,\nu_p)},
\end{equation}
where $\nu_p$ is the modulation frequency of the pump beam \cite{liu2013simultaneous,koh2007frequency}, which yields penetration depths of 100-300 nm in the temperature range of 30-300 K for all  metalattice samples.  We use reasonable estimates for thermal conductivity and heat capacity for the metalattice film. Since the length scale in all cases is much smaller than the spot sizes used in the measurement, the contribution from radial conduction is ignored and thermal transport is considered one-dimensional. 

Experimentally, the total laser power $\dot{Q}$ incident on filled metalattice samples is 65 mW at room temperature and is reduced to 6 mW at 30 K. For empty metalattice samples however the total power is taken as 40-50 mW at room temperature and reduced to 5 mW at 30 K. As a result, the temperature change due to steady-state heating is estimated to be
\begin{equation}
\Delta T = {(1 - \mathscr R) \dot{Q}}/{\sqrt{2\pi (w^2_0 + w^2_1) \Lambda_{\rm Si}^2}},
\end{equation}
where $\mathscr R$ is the reflectivity of aluminum, and $w_0$ and $w_1$ are the sizes of the pump and probe spots, respectively \cite{cahill2004analysis,braun2018steady}. In our setup, the temperature rise $\Delta T$ for all the samples in 30-300 K range is smaller than 1 K. This ensures the validity of both the single-temperature approximation and the condition of linear reflectance change.

\section{Supporting Information}
The \textit{Supporting Information} is available free of charge on the ACS Publications website at (...). The following supplementary content is provided: the experimental setup for the time-domain thermoreflectance measurements, the theoretical discussion of the ballistic length scales (self-diffusion and driven diffusion), the thermal conductivity calculation using the ballistic model with a benchmark application for silicon thin films, the calculation of the Rayleigh scattering efficiency, the reduction of the phonon group velocity in metalattices, and the discussion of phonon-isotope scattering effects.

\section{Acknowledgements}

The authors acknowledge financial support from the National Science Foundation Materials Research Science and Engineering Center for Nanoscale Science at Penn State University under grant NSF-DMR 1420620. 

The authors dedicate this contribution to the life and memory of John V. Badding.

\section{Table of Content Image}
\begin{figure}
    \centering\includegraphics[width=1\columnwidth]{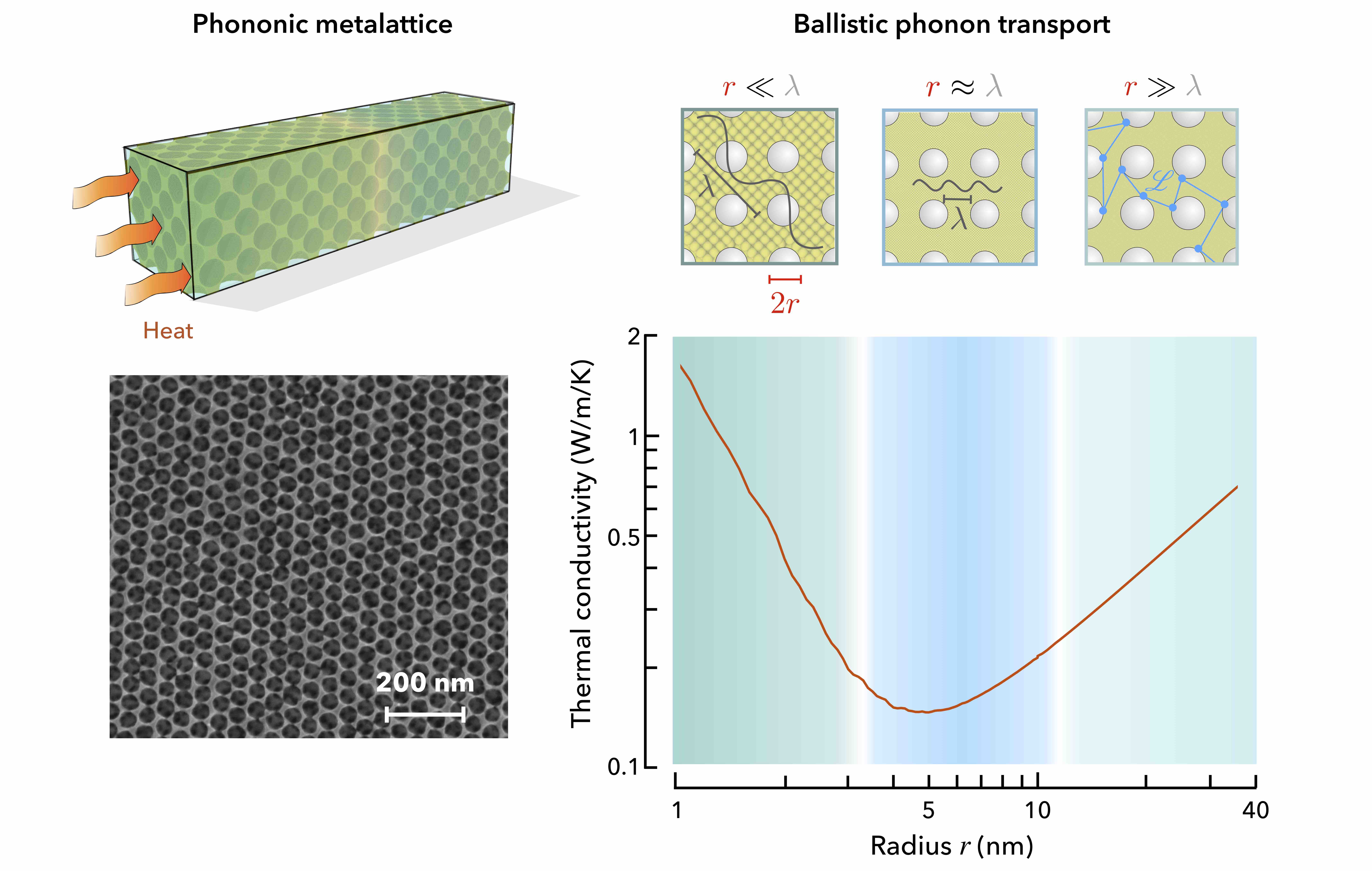}
    \label{fig:TOC}
\end{figure}

\providecommand{\latin}[1]{#1}
\makeatletter
\providecommand{\doi}
  {\begingroup\let\do\@makeother\dospecials
  \catcode`\{=1 \catcode`\}=2 \doi@aux}
\providecommand{\doi@aux}[1]{\endgroup\texttt{#1}}
\makeatother
\providecommand*\mcitethebibliography{\thebibliography}
\csname @ifundefined\endcsname{endmcitethebibliography}
  {\let\endmcitethebibliography\endthebibliography}{}


\end{document}